\documentstyle[11pt,newpasp,twoside,epsfig,here]{article}
\def\edcomment#1{\iffalse\marginpar{\raggedright\sl#1\/}\else\relax\fi}
\reversemarginpar

\begin{document}
\title{The XMM-Newton view of the radio-quiet neutron star 1E~1207.4--5209}

\author{A. De Luca, S. Mereghetti, P.A. Caraveo}
\affil{CNR/IASF ``G.Occhialini'', Via Bassini 15, I-20133 Milano, Italy}
\author{R.P. Mignani}
\affil{ESO, Karl Schwarzschild Str. 2, D-85740 Garching, Germany}
\author{W. Becker}
\affil{MPE, Giessenbachstrasse, Postfach 1312, D-85740 Garching, Germany}
\author{G.F. Bignami}
\affil{Universit\`a di Pavia, Pavia, Italy}

\begin{abstract}
We have observed the radio-quiet X-ray pulsar 1E~1207.4--5209 with the high 
throughput EPIC cameras onboard XMM-Newton. The spectrum of 
this peculiar source is characterized by two broad absorption features which 
present significant substructures and show a clear phase-dependence. We believe 
that these features represent a strong evidence for the presence of a magnetized 
atmosphere containing heavy elements.

\end{abstract}

\section{Introduction}
More than thirty years after the discovery of radio pulsars, the problem of 
understanding the properties of condensed matter in the interior of Neutron 
Stars (NSs) has remained largely unsolved. A promising way to obtain constraints 
on the equation of state of NSs is represented by the study of thermal radiation 
(peaking at X-ray energies) from their surfaces. The observed spectra are 
expected to show, as a characteristic signature of radiative transfer effects 
induced by the atmosphere surrounding the NS, several absorption features. 
These, if correctly deciphered, could unveil the NS physics. However, no such 
features were ever detected in the X-ray spectrum of a neutron star until recently, when Chandra and XMM-Newton observed 
the radio-quiet NS 1E~1207.4--5209.

This peculiar source (Bignami et al.1992; Mereghetti et al.1996), located close to the center of the high galactic latitude ($b \sim 10^{\circ}$) supernova remnant G296.5+10.0, was securely identified as a NS when fast pulsations (P$\sim$424 ms) were detected by Chandra (Zavlin et al.2000); a second Chandra observation led to the discovery of two broad 
absorption features in its X-ray spectrum (Sanwal et al. 2002).

Here we report on an XMM-Newton observation (30 ksec) performed on 2001, December. The high throughput of the EPIC instrument allowed for a deeper study of the phenomenology of 1E~1207.4--5209. See Mereghetti et al.(2002) for more details.

\section{The XMM/EPIC results}

We measured a period of 424.13084$\pm$0.00046 ms,  
implying a surprisingly low value of the 
period derivative, ${\dot P}$=(1.98$\pm$0.83)$\times$10$^{-14}$ s s$^{-1}$, in 
agreement with (but more accurate than) the Chandra estimate (Pavlov et al. 
2002). 


\begin{figure}[ht]
\centerline{\includegraphics[width=9cm,angle=-90]{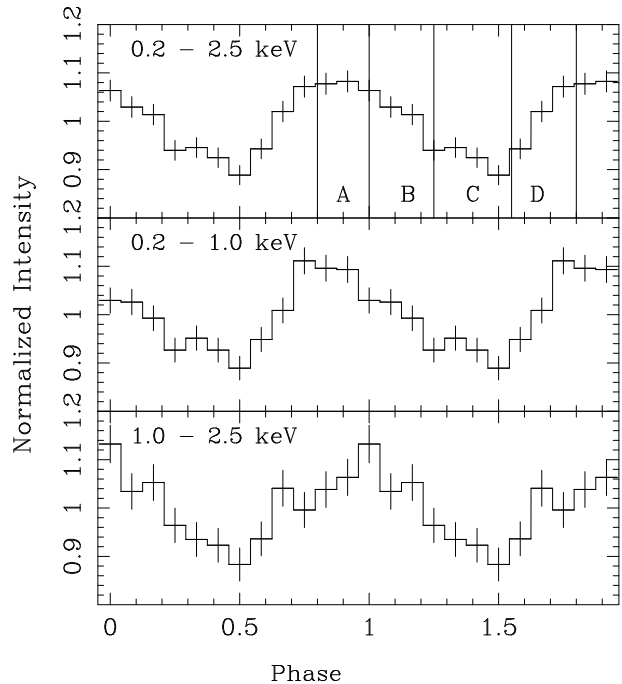}}
\caption{Energy-resolved light curves of 1E~1207.4--5209.}
\end{figure}

The pulse shape does not present any significant energy dependence (see Fig.1), 
at variance with the findings of Pavlov et al.(2002), who reported some evidence for a phase shift 
of 0.4$\div$0.6 between the energy bands 0.3-1.0 and 1.0-1.7 keV. 
If the latter result is confirmed, it might imply a time-variable light curve.

The spectrum of 
1E~1207.4-5209 is characterized by two broad absorption features at 0.7 and 1.4 
keV (Sanwal et al.2002). 
Figure 2 presents the results of the analysis of the EPIC pn spectrum
by showing the residuals obtained with different models. 


\begin{figure}[ht]
\centerline{\includegraphics[width=11cm,angle=-90]{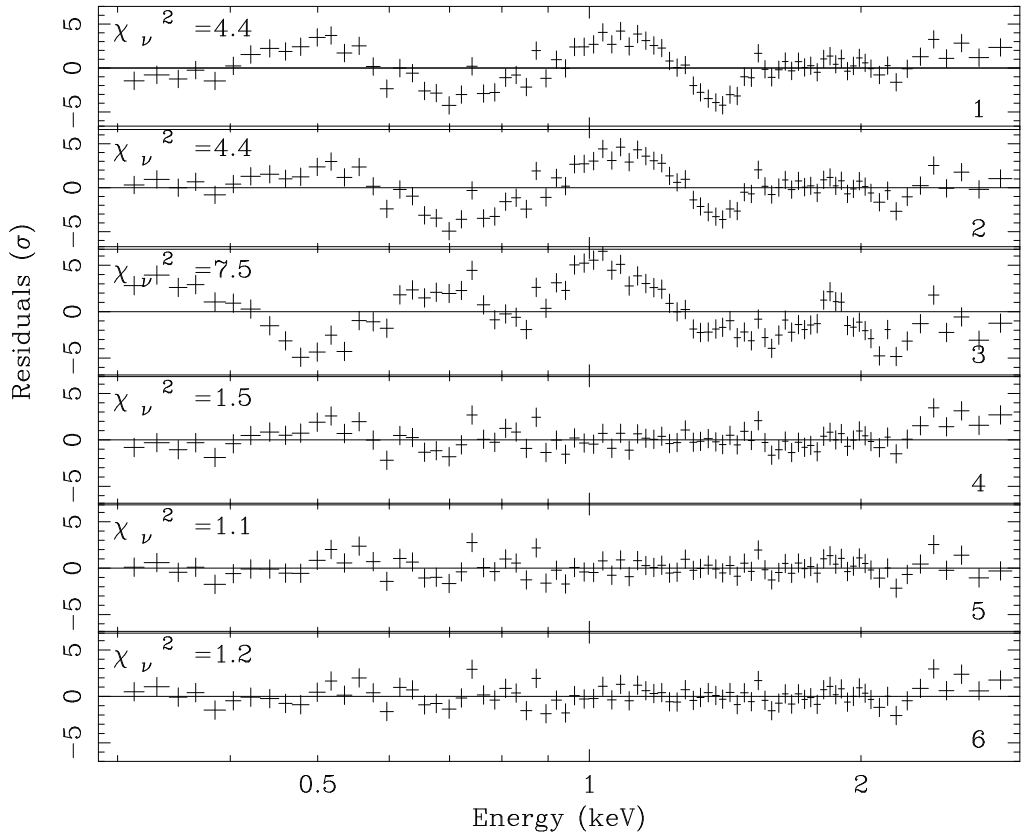}}
\caption{Results of phase-averaged spectroscopy. See text for details.}
\end{figure}

\begin{figure}[ht]
\centerline{\includegraphics[width=11cm,angle=-90]{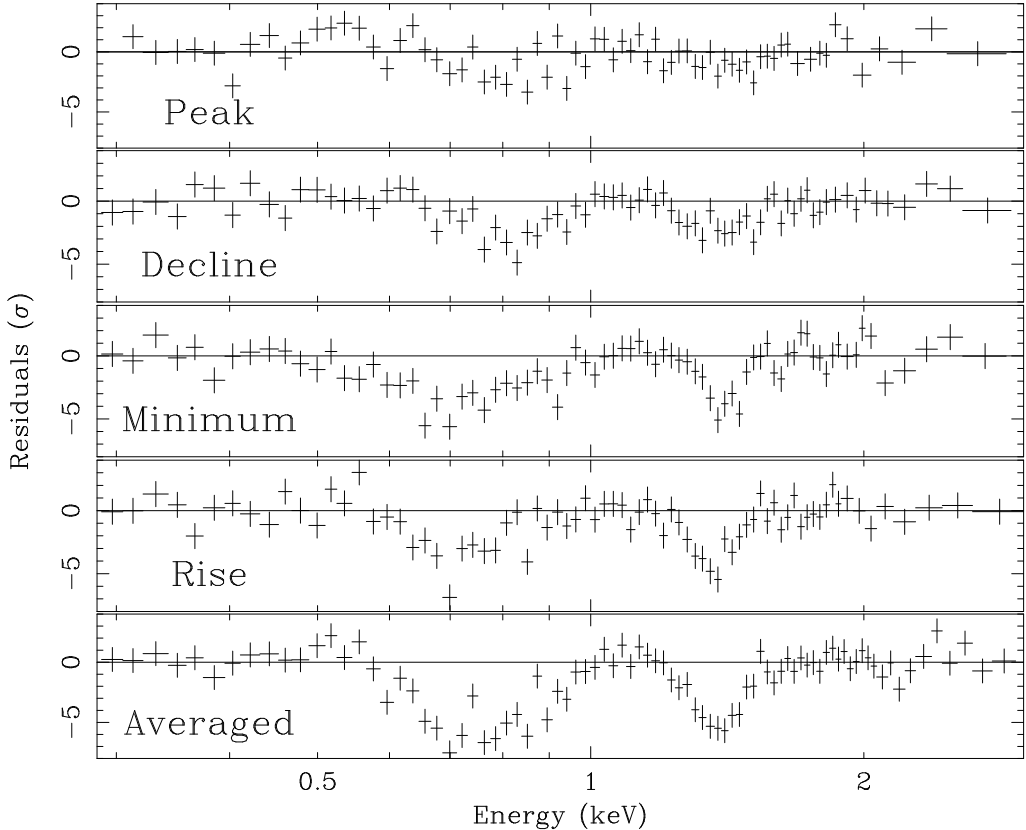}}
\caption{Results of phase-resolved spectroscopy. See text for details.}
\end{figure}

\begin{itemize}

\item Panel 1: a simple blackbody fits is clearly inadequate

\item Panel 2: the weakly-magnetized 
Hydrogen atmosphere model by Zavlin et al.(1996), although possibly 
inappropriate to the case of 1E~1207.4-5209 (inferred B$\sim 3 \times 10^{12}$ 
G), gives residuals very similar to the blackbody fit. 

\item Panel 3: models of NS atmospheres with solar 
abundance (G\"{a}nsicke  et al.2002) give 
unacceptable fits, since the predicted features are much narrower than 
the observed ones. The same is true for Fe composition.

\item Panel 4: acceptable fits required the inclusion of two absorption 
lines, that we modeled for simplicity as gaussian lines: here we give the 
residuals to a blackbody$+$lines fit. 

\item Panel 5: a further improvement was obtained by 
adding a power law: we believe that this reflects the shortcomings of 
other models alone to reproduce the low-energy part of this complex spectrum, 
rather than being evidence for a distinct non-thermal component. 

\item Panel 6: for completeness,  
the residuals to an hydrogen atmosphere$+$lines fit are shown here.

\end{itemize}

The plots in panels 4$\div$6 show that the broad feature at 0.7 keV is not well 
described by a single gaussian line. Significant substructures suggest 
that the feature may be due to the blending of several narrower lines. 

The EPIC pn data were divided in four subsets corresponding to the phase 
intervals shown in Fig.1. The resulting spectra provide striking evidence 
of pulse-phase variations of the spectral features. This is clearly shown in 
Fig.3: we have fitted a blackbody+power law model after excluding the energy bands corresponding to the features. The residuals on the overall energy range show that the line at 1.4 keV is more 
pronounced during the minimum and the rising part of the pulse profile, while it is 
almost unvisible during the pulse peak. The feature below 1 keV presents a 
significant variation in shape, possibly due to different contributions of 
several narrower lines during the various phase intervals. 

\section{Conclusions}
Our XMM-EPIC observation of 1E~1207.4-5209 has shown new, crucial details of the 
phenomenology of this puzzling source.

The timing analysis yielded a very low value for the period 
derivative. The resulting characteristic age, $\tau_{c}=(340\pm140)$ kyrs, turns out to be much higher than the age of the host supernova remnant,$\tau \sim$ 10 kyears. This discrepancy can be settled by supposing that the NS was born with a spin period very similar to the observed one.

Most important, the spectral analysis has revealed that the absorption features 
have significant substructure and are phase-dependent. This strongly supports an 
interpretation in terms  of atomic transitions in the neutron star atmosphere, 
different regions with different physical conditions being responsible for the 
emission visible at different phases. The actual chemical composition of the 
atmosphere is an open problem: Sanwal et al.(2002) proposed once ionized Helium in a superstrong magnetic field ($>10^{14}$ G, hardly conciliable with the value 
inferred from the pulsar spin parameters); we prefer an interpretation in terms 
of heavier elements in a more conventional ($\sim 10^{12}$ G) magnetic field. 
Indeed, Hailey \& Mori (2002) suggested that the lines could be produced by 
He-like Oxigen or Neon, predicting the presence of substructures in the broad 
features.

A second, very deep (250 ksec) XMM observation of 1E~1207.4-5209 has now been successfully performed (August 2002). The new data will allow for a more 
detailed analysis of phase-resolved spectra. This will yield important and maybe 
conclusive information on the atmosphere composition, hopefully providing firm 
constraints on the other neutron star parameters.

\end{document}